\documentclass[epsfig,12pt] {article}
\usepackage{graphics}
\usepackage{epsfig}
\usepackage{amssymb}
\usepackage{amsmath}
\usepackage{bm}
\usepackage{epsfig}

\begin{document}

\baselineskip .7cm

\author{ Navin Khaneja \thanks{To whom correspondence may be addressed. Email:navinkhaneja@gmail.com} \thanks{Systems and Control Engineering, IIT Bombay, Powai - 400076, India.}}

\vskip 4em

\title{\bf Chirp Mixing}

\maketitle

\vskip 3cm

\begin{center} {\bf Abstract} \end{center}
In this paper, we develop the theory of chirp mixing. The working principle is simple, given coupled homonuclear spins with offsets in range $[-B, B]$, we adiabatically sweep through the resonances. This achieves cross polarization between the $z$ magnetization of the coupled spins. We repeat this basic operation many times with a supercycle to achieve appropriate mixing time. When we sweep through the resonances, midway between the resonances of the coupled spin $I$ and $S$, the effective field seen by two spins is the same and hence they precess at same frequency around their effective fields. This means the coupling, which normally gets averaged due to the chemical shift difference is no more averaged for a short time and we get mixing. In this paper, we develop these basic ideas. By virtue of its design, the chirp mixing is much more broadband compared to state of the art methods. The proposed  methodology is demonstrated on $^{13}$C mixing in a sample of Alanine.  
\section{Introduction}

Total correlation spectroscopy (TOCSY) is a very important experiment in high resolution NMR spectroscopy. It performs cross-polarization between 
network of homonuclear spins with their offsets distributed in a given range. It provides total correlation between spins on backbone and side-chain of proteins and biomolecules,  which helps in assigning resonances \cite{cavanagh}. The work on mixing can be traced back to {\bf HOHAHA} by Ernst and coworkers \cite{hohaha, muller} on Hartmann-Hahn cross polarization. The seminal mixing sequences were narrow band. Since then there have been many developments in the field of broadband cross-polarization \cite{hohaha}-\cite{glaser}. These include phase-modulated irradiation schemes such as {\bf MLEV-17} \cite{bax1} and {\bf WALTZ-16}\cite{waltz1, waltz2}, {\bf IICT} \cite{bai}, {\bf NOIS} \cite{rao} and {\bf GD}\cite{drobny, glaser}. These made mixing broader. The bandwidth was further improved by {\bf DIPSI} and {\bf FLOPSY} family of pulse sequences developed by Shaka and coworkers \cite{dipsi, flopsy}. Recently method of multiple rotating frames have been used for design of mixing sequences \cite{paul}.

Given the coupled spin pair $I$ and $S$ with offsets $(\omega_I, \omega_S) = 2 \pi (\nu_I, \nu_S)$, the two spin Hamiltonian in a rotating frame is 
\begin{equation}
H = \omega_I I_z + \omega_S S_z + 2 \pi J I \cdot S.
\end{equation} If $\omega_I = \omega_S$, the coupling Hamiltonian $I \cdot S$ is preserved and transfers magnetization from one spin onto other, giving mixing. When $\omega_I - \omega_S \gg J$, the isotropic Hamiltonian $I \cdot S$ is truncated to $I_zS_z$, which does not transfer magnetization between spins. The working principle of the mixing sequences is to use rf-irradiation to eliminate/average the chemical shifts. Since the rf-Hamiltonian $\omega_1 (I_x+S_x)$ commutes with coupling, it only effects the chemical shifts and the goal is find a good sequence that averages the chemical shifts. While a cw irradiation is successful in removing the chemical shifts, one needs very large rf power to successfully eliminate large chemical shifts which can be in 10-20 kHz range for carbon mixing at high fields. Instead, suitable phase modulated sequences have proved to be more power efficient in effectively removing the chemical shits \cite{bax1, waltz1, waltz2, bai, rao, drobny, dipsi, flopsy, paul}. Nonetheless, these sequences do not maintain the coupling tensor perfectly. There is always a scaling of the chemical shift. The merit of a good mixing sequence is how broad range of frequency it can cover and achieve a good scaling factor.   

In this paper we introduce the theory of chirp mixing. The working principle is simple, given coupled homonuclear spins with offsets in range $[-B, B]$, we adiabatically sweep through the resonances. This achieves cross polarization between the $z$ magnetization of the coupled spins. This pulse sequence is repeated for appropriate mixing time. When we sweep through the resonances, midway between the resonances of the coupled spin $I$ and $S$, the effective field seen by two spins is the same and hence they precess at same frequency around their effective fields. This means the coupling, which normally gets averaged due to the chemical shift difference is no more averaged and we get mixing. We find the resulting mixing sequence is very broad because we can simply sweep through large range of resonances and all the action happens when we sweep midway between the two resonances. There is however a scaling of coupling which depends on the ratio of the chemical shift difference to rf-amplitude $\omega_1$. 

The paper is organized as follows. In section \ref{sec:theory}, we present the theory of chirp mixing. 
In section \ref{sec:results}, we present simulation and experimental results obtained using chirp mixing. We compare the mixing performance of chirp mixing with state of the art sequences like DIPSI and find chirp mixing is much more broader. We experimentally demonstrate chirp mixing on $^{13}$C mixing in an amino acid Alanine at high fields. Finally in section \ref{sec:conclusion}, we present conclusion and outlook.  

\section{Theory}
\label{sec:theory}
Let $$ \Omega_x = \left [ \begin{array}{ccc} 0 & 0 & 0 \\ 0 & 0 & -1 \\ 0 & 1 & 0 \end{array} \right ] , \ \Omega_y = \left [ \begin{array}{ccc} 0 & 0 & 1 \\ 0 & 0 & 0 \\ -1 & 0 & 0 \end{array} \right ], \ \Omega_z = \left [ \begin{array}{ccc} 0 & -1 & 0 \\ 1 & 0 & 0 \\ 0 & 0 & 0 \end{array} \right ]. $$ 
 denote generator of rotations around $x, y, z$ axis respectively. A x-rotation by flip angle $\theta$ is 
$ \exp( \theta \Omega_x)$. To fix ideas, we start be talking about single spin $\frac{1}{2}$.

\begin{figure}[htb!]
\begin{center}
\includegraphics [scale = .5]{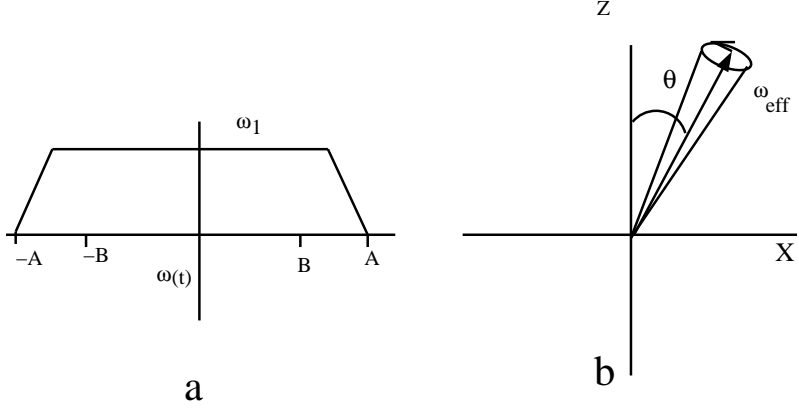} \\
\caption{Fig. a shows the amplitude $\omega_1$ of the chirp pulse as function of sweep frequency $\omega(t)$ as it is swept from $-A$ to $A$. Fig b shows how $Y(t)$ in Eq. (\ref{eq:phiframe}) follows the effective field as $\theta$ goes from $0$ to $\pi$. } \label{fig:sweep}
\end{center}
\end{figure}

Given the Bloch equations in rotating frame,

$$ \dot{X} = ( \omega_0 \Omega_z + \omega_1 \cos \phi \ \Omega_x +  \omega_1 \sin \phi \ \Omega_y ) X, $$ where $X$ is the magnetization vector and $\omega_0$ the offset. The chirp pulse has instantaneous frequency  $ \dot{\phi} = \omega = - A + at$ where $a$ is the sweep rate and phase  $\phi(t) = -At + \frac{at^2}{2}$. The
frequency $\omega$ is swept from $[-A, A]$, in time $T = \frac{2A}{a}$ with offsets $\omega_0$ in the range $[-B, B]$. See Fig. \ref{fig:sweep}a.

In the interaction frame of the chirp phase, $\phi(t)$, we have $Y(t) = \exp(-\phi(t) \Omega_z) X(t)$, evolve as  
\begin{equation}
\label{eq:phiframe}
 \dot{Y} = ( (\omega_0 - \omega) \Omega_z + \omega_1 \Omega_x ) Y = \omega_{\rm eff} \ (\cos \theta(t) \ \Omega_z + \sin \theta(t) \ \Omega_y) Y,
\end{equation}
where effective field strength $\omega_{\rm eff} = \sqrt{(\omega_0 - \omega(t))^2 + \omega_1^2}$ and $\tan \theta(t) = \frac{\omega_1}{\omega_0 - \omega(t)}$. See Fig. \ref{fig:sweep}b.

Now in interaction frame of $\theta$ where $Z = \exp(- \theta(t) \Omega_y) Y $, we have 

$$  \dot{Z}  = ( \omega_{\rm eff} \Omega_z -  \dot{\theta} \Omega_y) Z. $$

If $\omega_{\rm eff} \gg \dot{\theta} $, which is true when sweep rate and rf-field strength satisfy $a \ll \omega_1^2$, in the interaction frame of $W = \exp(- \int_0^t  \omega_{\rm eff} \ dt \ \Omega_z) Z $, we average $W(t)$ to $I$. Therefore the evolution of the Bloch equation for the chirp pulse takes the form

\begin{equation}
\label{eq:main}
X(t) = \exp(\phi(t) \ \Omega_z) \exp(\theta(t) \ \Omega_y)  \exp( \int_0^t \omega_{\rm eff}(t) \ dt  \Omega_z) X(0). 
\end{equation} 
Eq (\ref{eq:main}) is our main equation. 

We now understand how we can use chirp pulse for mixing. We consider coupled spin
homonuclear spin pair $I$ and $S$ coupled by a scalar coupling. In a rotating frame, the Hamiltonian of the system 
takes the form 

\begin{equation}
H(t) = \omega_I I_z + \omega_s S_z + \omega_1(t) (\cos \phi(t) F_x + \sin \phi(t) F_y) + 2 \pi J I \cdot S, 
\end{equation} where $\omega_I, \omega_S$ are chemical shifts of spin $I$ and $S$ and $F_x = I_x + S_x$, etc.
The chemical shifts are in the range $[-B, B]$. We now apply a chirp pulse that sweeps through the 
 this frequency range. $\omega_1$ and $\phi$ are the amplitude and phase of the chirp pulse. We proceed in the interaction frame of chemical shifts and rf-field and look at coupling Hamiltonian in this
frame. Then the coupling evolves as 

\begin{equation}
H_c(t) = 2 \pi J \ U'(t)(I \cdot S)U(t) , 
\end{equation} where $U(t)$ is the chemical shifts and rf-field propagator and from Eq (\ref{eq:main}), we have 

\begin{equation}
U(t) =  \exp(-i \phi(t) \ F_z)  \underbrace{\exp(-i (\ \theta_I(t) I_y +  \theta_S(t) S_y \ ))}_{U_\theta(t)}  \exp( -i \int_0^t \omega^I_{\rm eff}(t)I_z  +  \omega^S_{\rm eff}(t)S_z \ dt ).
\end{equation} Observe $U'(t)(I \cdot S)U(t) =$

\begin{equation}
\label{eq:average}
\exp( i \int_0^t \omega^I_{\rm eff}(t)I_z  +  \omega^S_{\rm eff}(t)S_z \ dt ) \ U'_\theta(t) (I \cdot S)  U_\theta(t) \ \exp( -i \int_0^t \omega^I_{\rm eff}(t)I_z  +  \omega^S_{\rm eff}(t)S_z \ dt ).
\end{equation}

\begin{figure}
\begin{center}
\includegraphics [scale = .5]{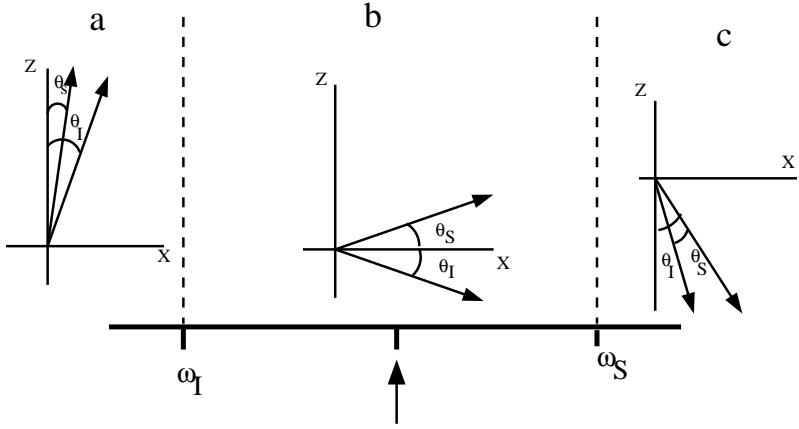}
\caption{Figure shows how $\theta_I$ and $\theta_S$ for coupled spins $I$ and $S$ change as we sweep through the resonances. When we are far left to both resonances (a), we have $\theta_I, \theta_S \sim 0$. When we sweep midway between two resonances (b)
$\theta_I, \theta_S \sim \frac{\pi}{2}$ (for $\omega_1 > \frac{\Delta}{2}$) and when we sweep far right to two resonances (c),  $\theta_I, \theta_S \sim \pi$.} \label{fig:resonances}
\end{center}
\end{figure}

We are interested in evaluating $\int_0^T H_c(\tau) d \tau$ where $T$ is duration of the chirp pulse. We first qualitatively understand this. As we sweep, when we are far left of the resonances of $I$ and $S$ (in region $a$ of fig.  \ref{fig:resonances}) , we have $ \theta_I, \theta_S \sim 0$,  $ U'_\theta(t) (I \cdot S)  U_\theta(t) \sim I \cdot S$. $  \omega^S_{\rm eff}(t) -  \omega^I_{\rm eff}(t) =  \omega^S_{\rm eff} -  \omega^I_{\rm eff} = \Delta $. This in Eq. (\ref{eq:average}) averages the planar part $I_xS_x + I_yS_y$ of $I \cdot S$.
Thus we donot get any coupling. When we sweep between the two resonances $\omega_I$ and $\omega_S$  (in region $b$ of fig.  \ref{fig:resonances}) , say we are at midpoint between them, then 
$  \omega^S_{\rm eff}(t) =  \omega^I_{\rm eff}(t)$, then we do not average the coupling and full coupling evolves. Regarding $ \theta_I, \theta_S$, lets say $\omega_1 >  \frac{\Delta}{2}$, then at midpoint between the two resonances,  $ \theta_I, \theta_S \sim \pi/2$ and therefore $ U'_\theta(t) (I \cdot S)  U_\theta(t) \sim I \cdot S$. Therefore, when we sweep midway through two resonances, we get coupling. Then as we sweep far right of the resonances of $I$ and $S$,  (in region $c$ of fig.  \ref{fig:resonances}) we have $ \theta_I, \theta_S \sim \pi$,  $ U'_\theta(t) (I \cdot S)  U_\theta(t) \sim I \cdot S$. $  \omega^S_{\rm eff}(t) -  \omega^I_{\rm eff}(t) = - \Delta $. This in Eq. (\ref{eq:average}) averages the planar part $I_xS_x + I_yS_y$ of $I \cdot S$.
Thus we do not get any coupling. In nutshell coupling evolution happens when we we sweep between the two resonances.

We can be more precise what happens midway between resonances when  $\omega_1 \sim \frac{\Delta}{2}$, then  its not strictly true that $ \theta_I, \theta_S \sim \pi/2$. Nonetheless, 

\begin{eqnarray}
\label{eq:couplingev}
\nonumber U'_\theta(t) (I \cdot S)  U_\theta(t) &=& I_yS_y + \cos (\theta_I - \theta_S) (I_xS_x + I_zS_z) + \sin  (\theta_I - \theta_S) (I_zS_x - I_xS_z) \\
&=& \cos^2(\frac{\theta_I - \theta_S}{2}) (I_xS_x + I_yS_y) + \dots 
\end{eqnarray} The coupling is attenuated by a factor  $\cos^2(\frac{\theta_I - \theta_S}{2}) = \frac{\omega_1^2}{\omega_1^2 + \frac{\Delta^2}{4}}$, at half way between resonances. Which is understandable, as when chemical shift difference between resonances is larger than rf-field amplitude we get 
less coupling. 

The effective Hamiltonian then takes the form

\begin{equation}
\label{eq:H1}
H_1 = \int_0^T H_c(\tau) d \tau = a I_zS_z + b (\cos \alpha (I_xS_x + I_yS_y) + \sin \alpha (I_xS_y - I_yS_x)) + \dots. 
\end{equation}

where $\dots$ are small double quantum terms (they rotate at sum of  $\omega^I_{\rm eff}$ and $\omega^S_{\rm eff}$ and average out) . The zero quantum Hamiltonian $ b (\cos \alpha (I_xS_x + I_yS_y) + \sin \alpha (I_xS_y - I_yS_x))$ transfers $I_z \rightarrow S_z$.

To build enough coupling, we repeat the process and sweep every $T$ units. We can evaluate the effective coupling Hamiltonian every $2T$ units of time.
Then after $2T$ time units effective coupling is 

$$ H_2 = H_1 + \underbrace{\exp( i \int_0^T \omega^I_{\rm eff}(t)I_z  +  \omega^S_{\rm eff}(t)S_z \ dt )}_{U_S} \ U'_\theta \ H_1 \ U_\theta \ \exp( -i \int_0^T \omega^I_{\rm eff}(t)I_z  +  \omega^S_{\rm eff}(t)S_z \ dt ), $$ where $U_\theta \sim \exp(-i \pi (I_y + S_y))$, i.e,

\begin{eqnarray}
\label{eq:H2}
H_2 &=& H_1 + U_S \ ( a I_zS_z + b (\cos \alpha (I_xS_x + I_yS_y) - \sin \alpha (I_xS_y - I_yS_x) + \dots ) \ U_S'\\
\nonumber &=& H_1 + ( a I_zS_z + b (\cos (\alpha + \beta)  (I_xS_x + I_yS_y) - \sin (\alpha + \beta) (I_xS_y - I_yS_x) + \dots) \\
\nonumber &=& ( a' I_zS_z + b' (\cos \alpha' (I_xS_x + I_yS_y) + \sin \alpha' (I_xS_y - I_yS_x) + \dots)
\end{eqnarray} where $\beta$ is a phase shift introduced by $U_S$.

$H_2$ is the effective coupling after time $2T$ in a frame of chemical shifts and rf-field. The unitary propagator after time $T$ in this frame is just

$$ U(T) =   U_{\theta} U_S \sim \exp(-i \pi (I_y + S_y)) U_S. $$ After time $2T$ the propagator is 

\begin{equation}
\label{eq:cancelation}
U^2(T) \sim \exp(-i \pi (I_y + S_y)) U_S \exp(-i \pi (I_y + S_y)) U_S \sim I.
\end{equation}

Therefore after $2T$, the chemical shift-rf propagator returns to identity and we produce only coupling evolution $\exp(-iH_2)$ and we now simple repeat this basic cycle for the right mixing time.

Note in practice, the adiabatic inversion produced by chirp is not perfect. Therefore we can write the propagator for say spin $I$ after time $T$ as 

$$ U^I = \exp(-i \theta_1 I_z)  \exp(-i (\pi - \epsilon) I_y)  \exp(-i \theta_2 I_z), $$ where when $\epsilon = 0$, we have perfect inversion.

The effect of this imperfect inversion is that errors can pile up and perfect cancellation as in Eq (\ref{eq:cancelation}) is not true.
However if after $T$ we advance the phase of chirp pulse by $\pi$ then during second half of $2T$ we produce the propagator

$$ \bar{U}^I = \exp(-i \pi I_z)  U^I  \exp(i \pi I_z) = \exp(-i \theta_1 I_z)  \exp(i (\pi - \epsilon) I_y)  \exp(-i \theta_2 I_z) $$

Then after $2T$ we produce, 

\begin{eqnarray*}
&& \exp(-i \theta_1 I_z)  \exp(-i (\pi - \epsilon) I_y)  \exp(-i \theta_2 I_z)\exp(-i \theta_1 I_z)  \exp(i (\pi - \epsilon) I_y)  \exp(-i \theta_2 I_z) \\ &=&  \exp(-i \theta_1 I_z)   \exp(i \epsilon I_y) \exp(i \theta_1 I_z)\exp(i \theta_2 I_z)  \exp(-i \epsilon I_y) \exp(-i \theta_2 I_z) \\
&\sim& \exp(-i 2 (\theta_1 + \theta_2)  \sin \frac{\epsilon}{2} ( \sin \frac{\epsilon}{2} I_z +   \cos \frac{\epsilon}{2} I_x ) ))
\end{eqnarray*} We still have error that is linear in $\epsilon$. If now we do a super cycle $  U^I  \bar{U}^I \bar{U}^I U^I = $
\begin{eqnarray}
\label{eq:supercycle}
 \nonumber && \exp(-i 2 (\theta_1 + \theta_2)  \sin(\frac{\epsilon}{2}) ( \sin \frac{\epsilon}{2} I_z -   \cos \frac{\epsilon}{2} I_x ) ))  \exp(-i 2 (\theta_1 + \theta_2)  \sin(\frac{\epsilon}{2}) \sin (\frac{\epsilon}{2} I_z +   \cos \frac{\epsilon}{2} I_x ) )) \\ &\sim&   \exp(-i 4 (\theta_1 + \theta_2)  \sin \frac{\epsilon}{2} \sin \frac{\epsilon}{2} I_z))
\end{eqnarray} The error is now quadratic in $\epsilon$. By doing a supercycle over 4 periods of $T$ with phase during 2nd and 3rd periods advanced by $\pi$, we minimize the problem due to imperfect inversion.

\section{Results}
\label{sec:results}

\subsection{Simulations}

How much coupling do we get when we sweep through the resonances. From Eq. (\ref{eq:couplingev}), for 
$\omega_d (t) = \omega_{eff}^I(t) - \omega_{eff}^S(t)$, we have from Eq. (\ref{eq:average}),
\begin{eqnarray}
\label{eq:effcoupling}
\nonumber H_c(t) &=& 2 \pi J \cos^2 \frac{\theta_I(t) - \theta_S(t)}{2} \{ \cos \int_0^t  \omega_d (\tau) d \tau \ (I_xS_x + I_yS_y)
+ \sin \int_0^t \omega_d (\tau) d \tau  \ (I_xS_y - I_yS_x) \} + \dots, \\
&=& c(t) \ (I_xS_x + I_yS_y) + d(t)  \ (I_xS_y - I_yS_x) + \dots 
\end{eqnarray} where $\dots$ are small double quantum terms, which we neglect.

\begin{figure}[htb!]
\begin{center}
\includegraphics [scale = .5]{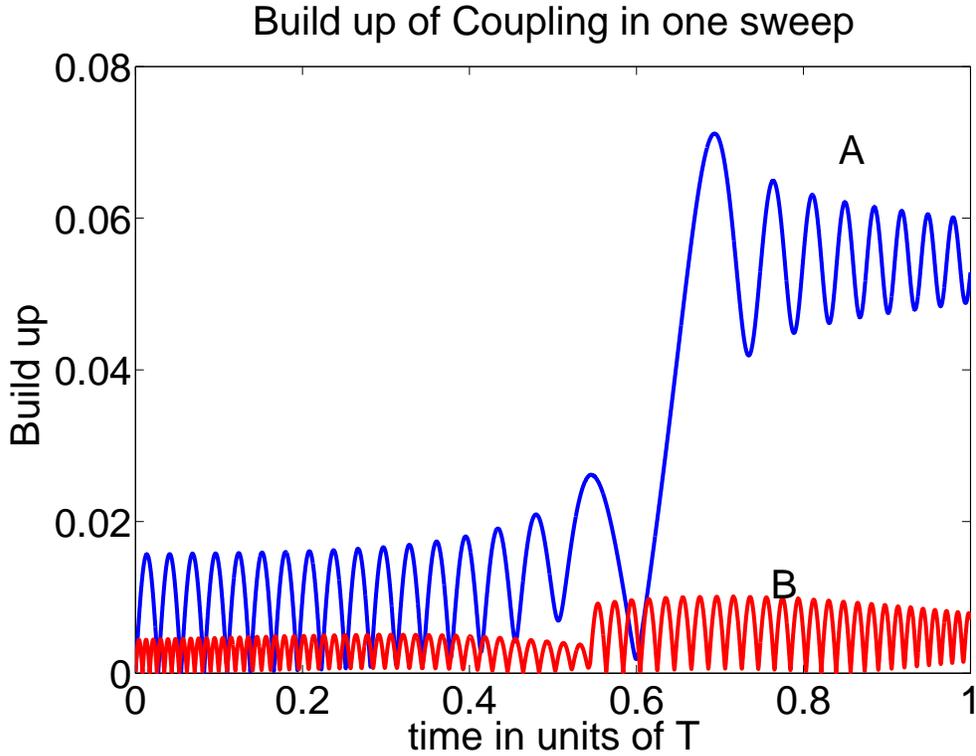}
\caption{A shows build up of zero quantum 
coupling Hamiltonian in one sweep of duration $T$. B shows build up of double quantum 
coupling Hamiltonian in one sweep of duration $T$.} \label{fig:zero-multiple}
\end{center}
\end{figure}

In Fig. \ref{fig:sweep}a and  \ref{fig:resonances}, we choose $A = 30$ kHz and $B = 20$ kHz. Let $\omega_I = - 5$ kHz and
$\omega_S = 20$ kHz, $J = 33$ Hz. We choose rf-field amplitude of $10$ kHz and a sweep rate $a = \frac{(2 \pi 10)^2}{16} (kHz)^2$, i.e.,
$a = \frac{\omega_1^2}{16}$ ( adiabaticity is well satisfied). The total time to sweep through $[-A, A]$ is then 
$T = 1.53$ ms. In Eq. \ref{eq:effcoupling}, we have for $\lambda \in \{I, S \}$, $ \omega_{eff}^\lambda(t) = \sqrt{\omega_1^2 + (\omega(t) - \omega_\lambda)^2} $
and $\tan \theta_\lambda(t) = \frac{\omega_1}{\omega_{\lambda} - \omega(t)}$. With these parameters, we plot in figure \ref{fig:zero-multiple} (curve A) , the build up of zero quantum 
coupling Hamiltonian given by 
\begin{equation}
\eta(t) = \sqrt{(\int_0^t c(\sigma) d \sigma)^2 +  (\int_0^t d(\sigma) d \sigma)^2},  
\end{equation}
in Eq. (\ref{eq:effcoupling}) in one period of chirp of duration $T$. We see in $T$ we accumulate a coupling of strength around $\sim .05$. To get a coupling of strength $\pi$ that transfers $I_z \rightarrow S_z$, we will need $\frac{\pi}{.05} \sim 63$ such periods, i.e. a duration of $96.4$ ms. For $J = 33 Hz$, this time is $\sim \frac{3}{J}$ units of time.
We also plot in  figure \ref{fig:zero-multiple} (curve B), build up of double quantum Hamiltonian. We find this is not much and hence $\dots$ terms in
 Eq. (\ref{eq:effcoupling}) are negligible. Further observe the build up of zero quantum 
coupling Hamiltonian happens when we sweep in the middle of the two resonances. In present simulation center of the resonances is at $12.5$ kHz and we sweep it at time $\frac{42.5}{60} T \sim .7 T$, which is where we see the build up in figure \ref{fig:zero-multiple} (curve A). Since double quantum terms in the Hamiltonian are rotated at sum of the effective frequencies $\omega_{eff}^\lambda$, they always average out as shown. 

\begin{figure}[htb!]
\begin{center}
\includegraphics [scale = .3]{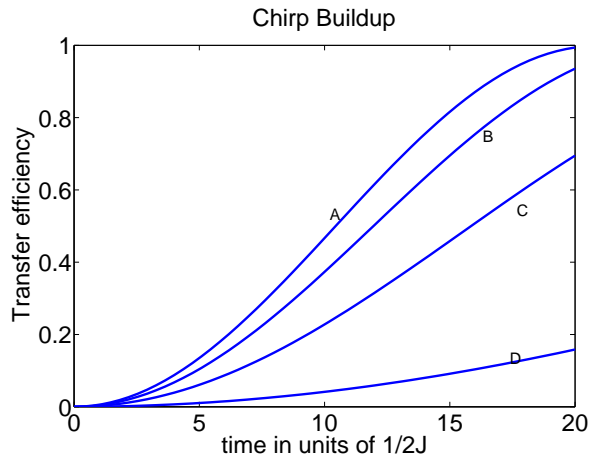}
\caption{The figure shows transfer of $I_z \rightarrow S_z$ as function of mixing time in units of $\frac{1}{2J}$. In above figure, A is $(\nu_I, \nu_S) = (-5, 10)$ kHz. B is  $(\nu_I, \nu_S) = (-5, 5)$ kHz. C is  $(\nu_I, \nu_S) = (-5, 20)$ kHz. D is  $(\nu_I, \nu_S) = (-5, 15)$ kHz. Rf-amplitude is $10$ kHz. } \label{fig:chirpbuild}
\end{center}
\end{figure}

In figure \ref{fig:zero-multiple}, we plotted build up of coupling $H_1$ in Eq. (\ref{eq:H1}) of one period $T$ of chirp. We can also plot  build up of coupling $H_2$ in Eq. (\ref{eq:H2}) in two periods of chirp. This is not simple sum of two $H_1$, there is a phase shift when we add effective coupling Hamiltonians in first $T$ period and second $T$ period as in  Eq. (\ref{eq:H2}). Because of this phase shift we find our transfer efficiency decreases and as shown in simulations below (see Fig. \ref{fig:chirpbuild}), it takes more than $\frac{3}{J}$ for the complete transfer in the pair $(\nu_I, \nu_S) = (-5, 20)$ kHz. We plot in Fig. \ref{fig:chirpbuild} for different resonance pairs, the build  up of transfer $I_z \rightarrow S_z$ as function of mixing time, when we do many cycles of chirp and use the supercycle as described in Eq. (\ref{eq:supercycle}).  Recall a zero quantum Hamiltonian $\pi ( \cos \gamma (I_xS_x + I_yS_y) + \sin \gamma (I_xS_y - I_yS_x) )$ transfer $(I_z -S_z) \rightarrow -(I_z - S_z)$ and hence transfers $I_z \rightarrow S_z$. Our chirp mixing builds a zero quantum Hamiltonian for the transfer.

\begin{figure}[htb!]
\begin{center}
\begin{tabular}{lr}
\includegraphics [scale = .3]{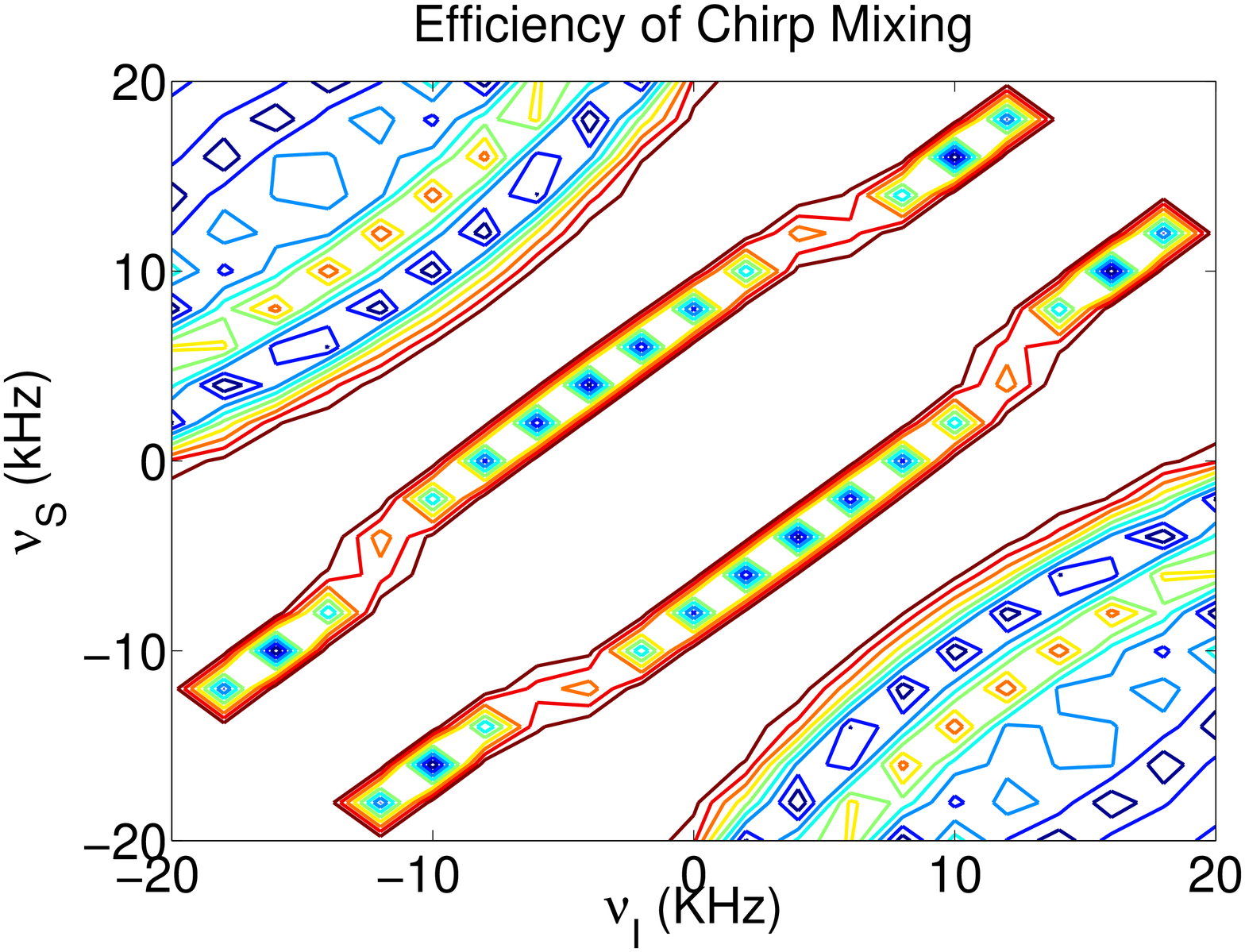} & \includegraphics [scale = .3]{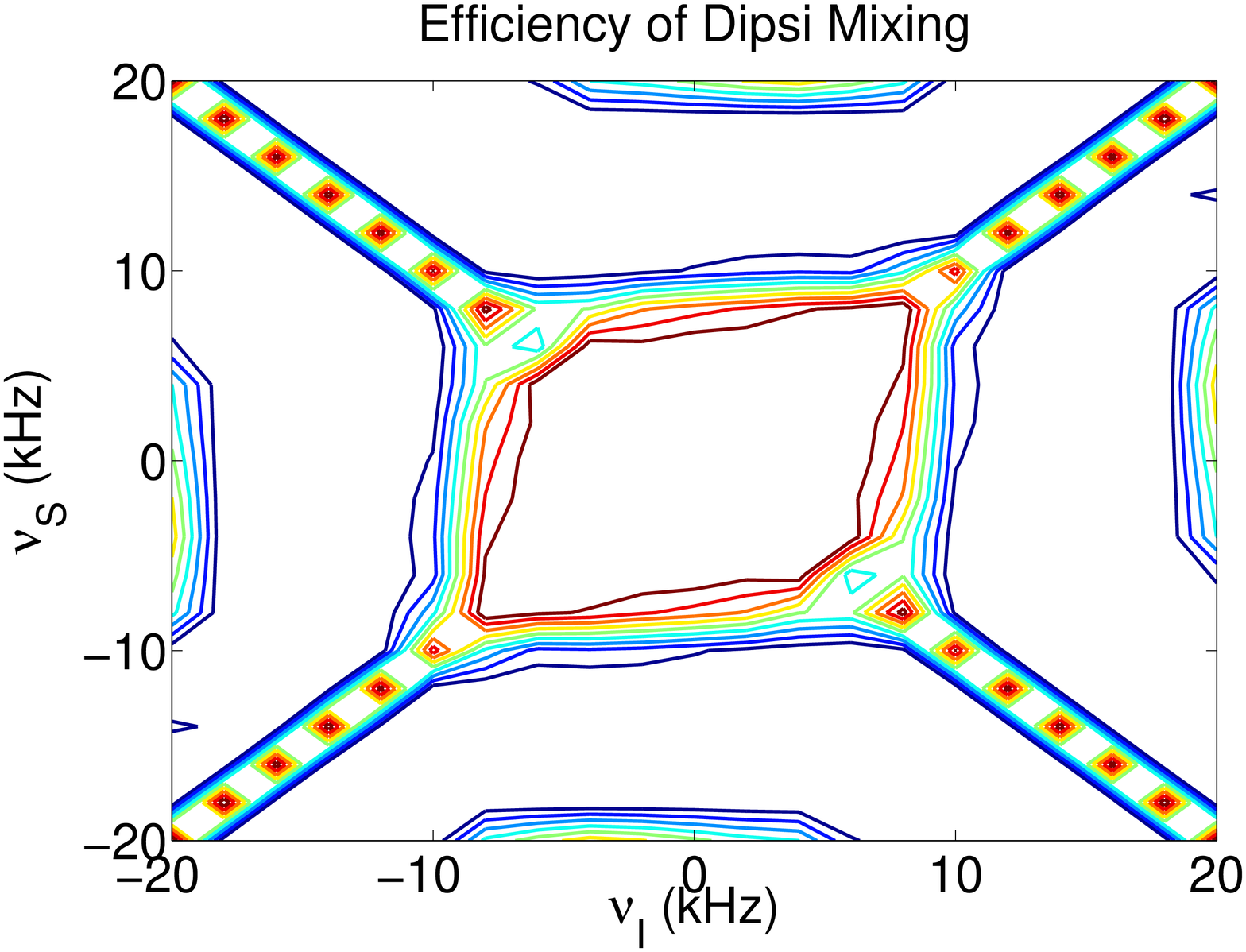}
\end{tabular}
\caption{The figure shows contour plot of transfer efficiency of the transfer $I_z \rightarrow S_z$ as function of offsets $\nu_I, \nu_S$. The transfer efficiency is the best achieved within time $\frac{10}{J}$. Left panel is for chirp mixing and right for DIPSI mixing. } \label{fig:contour}
\end{center}
\end{figure}

In Fig. \ref{fig:contour} (left panel), we plot maximum transfer efficiency within time $10/J$, as a function of resonance offsets $(\nu_I, \nu_S)$. We plot a contour plot, showing a large high intensity plateau around diagonal, which signifies chirp mixing is very broadband. We do the same for
state of the art DIPSI mixing (right panel) \cite{dipsi}. We find high intensity region around diagonal is much narrow. We find chirp mixing is much broader than DIPSI mixing.

\subsection{Experimental}
\begin{figure}[htb!]
\begin{center}
\includegraphics [scale = .25]{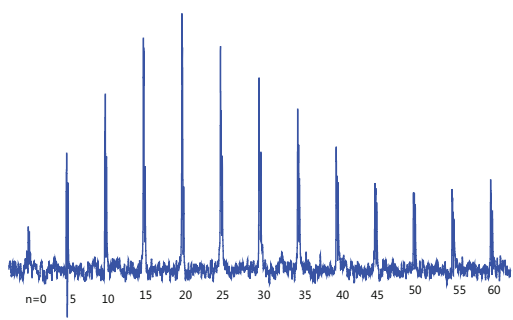}
\caption{The figure shows build up of transfer from C$_\beta$ to  C$_\alpha$ as we increment the mixing time of the chirp mixing by incrementing $n$, the number of supercycles of chirp mixing. The time of one supercycle is $T = 6.11$ ms. The maximum build up is at $n=20$, corresponding to a total mixing time of $122.2$ ms. } \label{fig:buildup}
\end{center}
\end{figure}

\begin{figure}[htb!]
\begin{center}
\includegraphics [scale = .55]{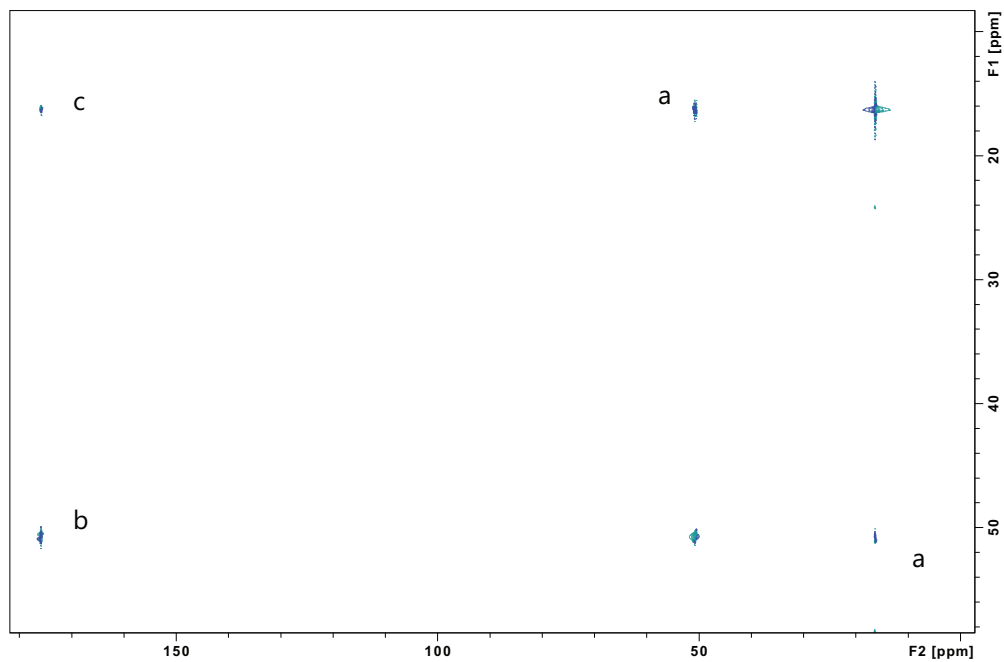} 
\caption{The figure shows a 2D $^{13}$C spectra obtained using chirp mixing on a molecule of Alanine.  We can see all C$_\beta$-C$_\alpha$ (a) ,  C$_\alpha$-CO (b) and C$_\beta$-CO (c) cross peaks. } \label{fig:chirpsydata}
\end{center}
\end{figure}

The chirp mixing pulse sequence was applied for $^{13}$C mixing in uniformly labelled amino acid Alanine. The molecule has three $^{13}$C resonances of C$_\beta$, C$_\alpha$ and CO at
$16, 50, 177$ ppm respectively with coupling $J_{C_\beta-C_\alpha}= 34$ Hz and $J_{CO-C_\alpha}= 55$ Hz. All experiments were performed at magnetic field corresponding to 750 MHz proton frequency. The chirp mixing pulse sequence used for the experiments is the one defined in simulation section. We put the carrier half way between the resonances at 100 ppm. To find a good mixing time for the experiment, we first do a 1D experiment, where by use of gradients and selective pulses, the magnetization on C$_\alpha$ and CO is dephased. Then $n$ supercycles of chirp mixing are applied (each supercycle takes time $T = 6.11$ ms) and we observe build up of magnetization on C$_\alpha$ starting from magnetization on C$_\beta$. $n$ is incremented in units of $5$ and we plot the build up in Fig. \ref{fig:buildup}. We find we get best transfer at $n=20$. This correspond to a mixing time of $122$ ms. With this mixing time we do a 2D experiment, results of which are shown in Fig. \ref{fig:chirpsydata}. The carrier is at $100$ ppm. We collect 1024 points in indirect and 4096 points in the direct dimension. We use TPPI for data acquisition. During mixing proton is decoupled with a decoupling pulse sequence. We can see C$_\beta$-C$_\alpha$ (a) ,  C$_\alpha$-CO (b) and C$_\beta$-CO cross peaks in Fig. \ref{fig:chirpsydata}.

\section{Conclusion}\label{sec:conclusion}
In this paper, we developed the theory of chirp mixing. The pulse sequence for chirp mixing is simple, given coupled homonuclear spins with offsets in range $[-B, B]$, we adiabatically sweep through the resonances. This achieves cross polarization between the $z$ magnetization of the coupled spins. We repeat this adiabatic sweep for appropriate mixing time, with a supercycle. When we sweep midway between the resonances of the coupled spin $I$ and $S$, the effective field seen by two spins is the same and hence they precess at same frequency around their effective fields. This means the coupling, which normally gets averaged due to the chemical shift difference is no more averaged and we get mixing. By virtue of its design, the chirp mixing is much more broadband compared to state of the art methods. We compared the performance of chirp mixing to DIPSI mixing. We find for a rf-field amplitude of $10$ kHz, DIPSI is only effective in a offset range $[-10, 10]$ kHz, while chirp mixing easily covers the range $[-20, 20]$ kHz and beyond. This range can be further extended, by increasing the sweep width. 

The proposed  methodology was demonstrated on $^{13}$C mixing in a sample of Alanine. We anticipate the proposed methods will soon find applications in broadband $^{13}$C  mixing in protein NMR spectroscopy at high fields. At field strengths of $1$ GHz the  $^{13}$C spectrum is spread over a $45$ kHz range, which can be easily covered with chirp mixing with moderate rf-field strengths of $10$ kHz.

\end{document}